\documentclass[10pt,fleqn,a4paper,twoside]{article}
\usepackage{eurodiname2026, bm}
\usepackage{amsfonts}
\usepackage[fleqn]{amsmath}
\def\shortauthor{K. Paldanius, G.S. Lima and W.M. Bessa}
\def\shorttitle{Influence of Radial Basis Activation Functions on Intelligent Controller for Robotic Manipulators}

\begin{document}
\fphead
\hspace*{-2.5mm}\begin{tabular}{||p{\textwidth}}
\begin{center}
\vspace{-4mm}
\title{EURODINAME-2026-9024036\\
INFLUENCE OF RADIAL BASIS ACTIVATION FUNCTIONS ON INTELLIGENT CONTROLLER FOR ROBOTIC MANIPULATORS} 
\end{center}
\authors{Kimmo Paldanius} \\
\authors{Gabriel Da Silva Lima} \\
\authors{Wallace Moreira Bessa} \\
\institution{Smart Systems Lab, Department of Mechanical Engineering, University of Turku, 20520, Finland.} \\ 
\institution{kkpald@utu.fi, gdasil@utu.fi, wmobes@utu.fi} \\
\\
\abstract{\textbf{Abstract.}
This paper presents an intelligent control framework for trajectory tracking of robotic manipulators using radial basis function (RBF) neural networks for online disturbance estimation. The proposed control structure combines model-based nonlinear control with an adaptive neural approximator that compensates for parametric uncertainties, friction, and unmodeled dynamics. A Lyapunov-based adaptation law with projection guarantees boundedness of the closed-loop signals and convergence of the tracking error to a compact region. The primary objective of this work is to investigate how the choice of activation function within the RBF network influences transient behavior, steady-state accuracy, and control smoothness. The controller is implemented on a robotic manipulator. Experimental results demonstrate that although stability is preserved for all kernels, activation function selection significantly affects adaptation dynamics and practical tracking performance. These findings demonstrate that activation function selection acts as a structural design parameter in intelligent control, directly shaping adaptation dynamics and practical closed-loop performance.}
\\
\\
\\
\end{tabular}

\section{INTRODUCTION}

Trajectory tracking of robotic manipulators is commonly addressed using model-based nonlinear control methods such as feedback linearization~\citep{ccari2024robust, liu2025neural}. By exploiting the known system dynamics, these approaches enable systematic cancellation of nonlinearities and precise reference tracking under nominal conditions. In practice, however, accurate tracking can be degraded by parametric uncertainty, friction, unmodeled dynamics, and external disturbances. To mitigate these effects, adaptive and learning-based extensions are often incorporated to enhance robustness while preserving the structure of model-based control~\citep{sveen2025decentralised, abu2026model}. Adaptive control techniques typically address structured parametric uncertainties through online parameter estimation mechanisms, with stability established using tools such as Lyapunov analysis or related adaptive frameworks, thereby ensuring boundedness of the closed-loop signals despite unknown system parameters~\citep{Ioannou2006, venanzi2016review, zhang2017review}. In contrast, learning-based approaches, particularly neural-network-based disturbance estimators, aim to approximate unmodeled or unstructured dynamics without requiring explicit parametric representations~\citep{brunke2022safe, zeng2025learning}. Owing to their universal approximation capability, neural networks are well suited to compensate for complex nonlinear effects such as friction and residual coupling dynamics.


This paper adopts a hybrid perspective that integrates feedback linearization with an online neural disturbance estimator based on radial basis function (RBF) networks. The neural weights are updated through a Lyapunov-based adaptation law, ensuring boundedness of all closed-loop signals and convergence of the tracking error to a neighborhood of the origin while exploiting the approximation capability of neural networks.


The primary focus of this work is the influence of the RBF activation (basis) function on closed-loop behavior. Although Lyapunov stability holds for a broad class of bounded activation functions, performance characteristics, such as transient response, steady-state accuracy, and control smoothness, can vary significantly with the kernel shape. While Gaussian kernels are commonly adopted by default in RBF-based controllers~\citep{niu2023adaptive, li2026rbfnn}, this study systematically investigates alternative activation functions, namely Laplacian and inverse multiquadratic kernels, under identical controller tuning and experimental conditions.

\begin{figure}[ht]
    \centering
    \includegraphics[width=0.4\linewidth]{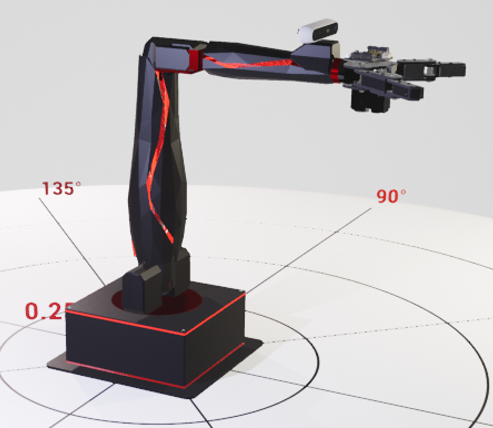}
    \caption{Quanser QArm (digital twin environment).}
    \label{fig:qarm}
\end{figure}

The approach is implemented on the Quanser QArm robotic manipulator, Fig.~\ref{fig:qarm}, and performance is evaluated using tracking error metrics and control effort. By isolating the activation function as the only varying element in the closed loop, the study demonstrates that kernel properties, such as locality, smoothness, and support, directly influence adaptation dynamics and practical tracking performance. These results provide design-oriented insight into activation function selection for neural-network-assisted feedback linearization.


The remainder of the paper is organized as follows. Section 2 develops the intelligent controller and its stability analysis, Section 3 discusses the activation functions, Section 4 presents the experimental results, and Section 5 provides the concluding remarks.


\section{INTELLIGENT CONTROLLER}

The equations of motion for an $n$-DOF robotic manipulator can be written as:
\begin{equation}\label{eq:modeln}
\bm{M}(\bm{q}) \ddot{\bm{q}} + \bm{k} (\bm{q},\dot{\bm{q}}) = \bm{h} (\bm{q},\dot{\bm{q}}) + \bm{u}
\end{equation}
where $\bm{q} \in \mathbb{R}^{n \times 1}$ is the vector of generalized coordinates, $\bm{u} \in \mathbb{R}^{n \times 1}$ the control input vector, $\bm{M}(\bm{q}) \in \mathbb{R}^{n\times n}$ is the inertia matrix, $\bm{k}(\bm{q},\dot{\bm{q}}) \in \mathbb{R}^{n \times 1}$ represents the centrifugal and Coriolis forces, and $\bm{h}(\bm{q},\dot{\bm{q}}) \in \mathbb{R}^{n \times 1}$ denotes the vector of external forces acting on the system.

For control purposes, the model equation~(\ref{eq:modeln}) can be rewritten as:
\begin{equation}\label{eq:model}
    \ddot{\bm{q}} = \hat{\bm{f}} + \hat{\bm{M}}^{-1}\bm{u} + \bm{d}
\end{equation}
with $\hat{\bm{M}}$ and $\hat{\bm{f}}$ being estimates of $\bm{M}$ and $\bm{f} = \bm{M}^{-1}(\bm{h} - \bm{k})$, respectively, and $\bm{d}$ representing the total uncertainty related to the dynamic model.

A control law based on the feedback linearization approach~\citep{slotine1991applied} is given by 
\begin{equation}\label{eq:control-law}
    \bm{u} = \hat{\bm{M}}(- \hat{\bm{f}} - \hat{\bm{d}} + \ddot{\bm{q}}_d - \bm{\Lambda} \dot{\tilde{\bm{q}}} - \bm{\Lambda} \bm{s})
\end{equation}
where $\bm{\Lambda} \in \mathbb{R}^{n \times n}$ is a diagonal matrix of strictly positive entries $\lambda_i$, $\bm{s} = \dot{\tilde{\bm{q}}} + \bm{\Lambda} \tilde{\bm{q}}$ is the combined error with $\tilde{\bm{q}} = \bm{q} - \bm{q}_d$ being the tracking error.

By applying the control~(\ref{eq:control-law}) into the Eq.~(\ref{eq:model}), the closed-loop dynamics becomes
\begin{equation}\label{eq:closed-loop}
    \dot{\bm{s}} + \bm{\Lambda} \bm{s} = \bm{d} - \hat{\bm{d}}
\end{equation}

From the Equation~(\ref{eq:closed-loop}), it can be verified that in the ideal case, when $\bm{d} = \hat{\bm{d}}$, the combined error $\bm{s}$ and, consequently, the tracking error $\tilde{\bm{q}}$ exponentially converge to zero. If it is not the case, closed-loop dynamics is driven by the approximation error. Furthermore, it also suggests that the combined error $\bm{s}$ represents a reasonable metric for tracking success and can be used to help calculate the estimate $\hat{\bm{d}}$.

\begin{figure}[t]
    \centering
    \includegraphics[width=0.4\linewidth]{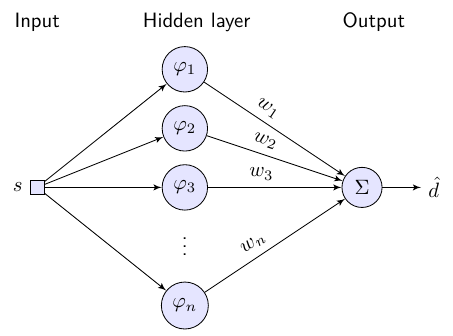}
    \caption{Architecture of the RBF neural network.}
    \label{fig:ann}
\end{figure}

Let us now propose that the components of the disturbance approximator $\hat{\bm{d}}$ is given by a RBF neural network, depicted in Fig.~\ref{fig:ann}:
\begin{equation}\label{eq:d_hat}
    \hat{d}_i = \bm{w}_i^\top \bm{\varphi}_i(s_i)
\end{equation}
where $\bm{w}_i = [w_{i,1}, w_{i,2}, \ldots, w_{i,n_i}]^\top$ is the vector of weights, $\bm{\varphi}_i = [\varphi_{i,1}, \varphi_{i,2}, \ldots, \varphi_{i,n_i}]^\top$ stands for the activation functions $\varphi_{i,j}$, with $i = 1, 2$ and $j = 1, \ldots, n_i$, and $n_i$ being the number of neurons in the hidden layer of the corresponding $i$th component of $\hat{\bm{d}}$.

Relying on the assumption that artificial neural networks can perform universal approximation~\citep{scarselli1998universal} with an arbitrary precision $\varepsilon_i$, it follows that $d_i = \hat{d}^\ast_i + \varepsilon_i$, where $\hat{d}^\ast_i$ is the output related with the optimal weight vector $\bm{w}_i^\ast$.

The boundedness and convergence properties of the closed-loop signals in the presence of modeling inaccuracies, equation~(\ref{eq:closed-loop}), can now be proved by means of a Lyapunov-like stability analysis.  Thus, let a positive-definite function $V_i$ be defined for each component of $\bm{s}$:
\begin{equation}
    V_i(t) = \frac{1}{2} s_i^2 + \frac{1}{2\eta_i} \bm{\delta}_i^\top \bm{\delta}_i
    \label{eq:lyap}
\end{equation}
with $\eta_i$ being a strictly positive constant and $\bm{\delta}_i = \bm{w}_i - \bm{w}_i^\ast$.

Since $\dot{\bm{\delta}}_i = \dot{\bm{w}}_i$, the time derivative of $V_i$ is
\begin{equation}
\begin{split}
    \dot{V}_i(t) & = s_i \dot{s}_i + \eta_i^{-1} \bm{\delta}_i^\top \dot{\bm{w}}_i =  [-\lambda_i s_i + d_i - \hat{d}_i]s_i + \eta_i^{-1} \bm{\delta}_i^\top \dot{\bm{w}}_i\\
                 & = -[\lambda_i s_i - (\hat{d}_i^\ast + \varepsilon_i - \hat{d}_i) ]s_i + \eta_i^{-1} \bm{\delta}_i^\top \dot{\bm{w}}_i\\
                 & = -[\lambda_i s_i - \varepsilon_i + \bm{\delta}_i^\top \bm{\varphi}_i ]s_i + \eta_i^{-1} \bm{\delta}_i^\top \dot{\bm{w}}_i\\
                 & = -[\lambda_i s_i - \varepsilon_i ]s_i + \eta_i^{-1} \bm{\delta}_i^\top [\dot{\bm{w}}_i - \eta_i s_i\bm{\varphi}_i]
    \label{eq:lyapdot}
\end{split}
\end{equation}

Hence, by updating $\bm{w}_i$ according to $\dot{\bm{w}}_i = \eta_i s_i \bm{\varphi}_i$, the time derivative becomes $ \dot{V}_i(t) = -[\lambda_i s_i - \varepsilon_i ]s_i \leq - [\lambda_i \vert s_i\vert - \epsilon_i ]\vert s_i\vert$, i.e.\ negative definite when $\vert s_i\vert > \epsilon_i / \lambda_i$. It means that the bounds of $\bm{w}_i$ cannot be ensured with $\dot{\bm{w}}_i = \eta_i s_i \bm{\varphi}_i$ when $\vert s_i\vert \leq \epsilon_i / \lambda_i$. Fortunately, the projection algorithm~\citep{Ioannou2006} can be evoked to guarantee that $\bm{w}_i$ will remain within a convex region $\mathcal{W}_i = \{\bm{w}_i \in \mathbb{R}^n : \bm{w}_i ^\top \bm{w}_i \leq \mu_i^2 \}$:
\begin{equation}
    \dot{\bm{w}}_i = \left\lbrace \begin{array}{cl}
        \eta_i s_i \bm{\varphi}_i & \text{if}~\Vert \bm{w}_i \Vert_2 < \mu_i ~\text{or}  \\ 
         & \text{if}~\Vert \bm{w}_i \Vert_2 = \mu_i ~\text{and}~ \eta_i s_i \bm{w}_i^\top \bm{\varphi}_i < 0 \\
         \left( \bm{I} - \frac{\bm{w}_i \bm{w}_i^\top}{\bm{w}_i^\top \bm{w}_i} \right)\eta_i s_i \bm{\varphi}_i  & \text{otherwise}
    \end{array} \right.
    \label{eq:projection}
\end{equation}
where $\mu_i$ is the desired upper bound of $\Vert \bm{w}_i \Vert_2$.

Since $\Vert \bm{w}_i (0) \Vert_2 \leq \mu_i$, it follows that $|s_i| \leq \epsilon_i/\lambda_i$ and $\Vert \bm{w}_i (t) \Vert_2 \leq \mu_i$ as $t \rightarrow \infty$. Then, recalling that $s_i = \dot{\tilde{q}}_i + \lambda_1 \tilde{q}_i$, it follows that the proposed controller ensures the exponential convergence of the tracking error to the closed region $\mathcal{Q} = \{ (\tilde{\bm{q}}, \dot{\tilde{\bm{q}}}) \in \mathbb{R}
^{4} : |\tilde{q}_i| \leq \epsilon_i \lambda_i^{-2}, |\dot{\tilde{q}}_i| \leq 2\epsilon_i \lambda_i^{-1}\}$~\citep{da2023accurate}.

\section{ACTIVATION FUNCTIONS FOR DISTURBANCE ESTIMATION}


Although Lyapunov stability is guaranteed for any bounded activation function under the proposed projection-based adaptation law, the kernel structure influences the approximation characteristics and, consequently, the evolution of the adaptive weights. Thus, the activation function should be viewed not merely as an implementation detail, but as a design parameter that shapes both transient and steady-state closed-loop behavior.


In the proposed framework, the lumped uncertainty term $\bm{d}$ in~(\ref{eq:model}) is compensated online by the neural approximator defined in~(\ref{eq:d_hat}), where the activation vector determines the mapping from the combined error  $s_i$ to the disturbance estimate $\hat d_i$. While the stability analysis in Section~2 ensures boundedness of all closed-loop signals under the projection mechanism~(\ref{eq:projection}), tracking performance ultimately depends on the approximation accuracy $d_i \approx \hat d_i$, which is directly affected by the choice of activation function.


Accordingly, this section examines how the kernel shape influences closed-loop performance. To enable a fair comparison, the control structure~(\ref{eq:control-law})–(\ref{eq:d_hat}), controller gains $\lambda_i$ and $\eta_i$, number of neurons $n_i$, and projection bounds $\mu_i$ are kept identical in all cases; only the activation functions are varied. Specifically, three radial basis kernels are considered: Gaussian (baseline), inverse multiquadratic, and Laplacian.




\subsection{Radial basis kernels}

Each neuron is defined from a normalized distance
\begin{equation}
    \rho_{i,j} = \frac{\bar s_i - c_{i,j}}{\sigma_i},
    \label{eq:rho}
\end{equation}
where $c_{i,j}\in\mathbb{R}$ is the center of the $j$th basis function and $\sigma_i>0$ is a width parameter. The $j$th
activation is then defined as
\begin{equation}
    \varphi_{i,j}(\bar s_i) = \kappa(\rho_{i,j}),
    \label{eq:phi_kernel}
\end{equation}
where $\kappa(\cdot)$ is one of the following kernels:

\begin{itemize}
\item \textbf{Gaussian (baseline):}
\begin{equation}
    \kappa_{G}(\rho) = \exp(-\rho^2).
    \label{eq:k_gaussian}
\end{equation}

\item \textbf{Inverse multiquadratic:}
\begin{equation}
    \kappa_{IMQ}(\rho) = \frac{1}{\sqrt{1+\rho^2}}.
    \label{eq:k_imq}
\end{equation}

\item \textbf{Laplacian:}
\begin{equation}
    \kappa_{L}(\rho) = \exp(-|\rho|).
    \label{eq:k_laplacian}
\end{equation}
\end{itemize}

The Gaussian and Laplacian kernels are localized and decay away from their centers. The inverse multiquadratic kernel has global support.
$|\rho|$. These differences motivate the comparative analysis presented in Section~4.

\subsection{Kernel parameterization and comparison protocol}

For all kernels, the same number of neurons $n_i$ is employed. The centers are chosen uniformly over the normalized
interval $\bar s_i \in [-1,1]$:
\begin{equation}
    c_{i,j} = -1 + \frac{2(j-1)}{n_i-1}, \qquad j=1,\ldots,n_i.
    \label{eq:centers_uniform}
\end{equation}
A common width is selected as
\begin{equation}
    \sigma_i = \gamma\,\frac{2}{n_i-1},
    \label{eq:sigma_rule}
\end{equation}
where $\gamma>0$ is a tuning factor, kept identical for all kernels.

In summary, the following elements are kept fixed across all tests: controller gains $\lambda_i$, learning rates
$\eta_i$, projection bounds $\mu_i$, number of neurons $n_i$, desired trajectories $\bm{q}_d(t)$, and all
simulation/experimental conditions. Consequently, any performance differences reported in Section~5 are attributed to the
activation kernel selection.

\section{RESULTS}

This section presents the experimental evaluation of the proposed intelligent controller with different radial basis activation functions. The goal is to assess how the kernel choice influences transient response, steady-state accuracy, and control effort. The control parameters were set as follows: $\lambda_i = 30$, $\eta_i = 10$, $\mu_i = 100$, and $n_i = 6$.


\subsection{Experimental Setup}


A single DOF manipulator was considered here. For this case, the generalized coordinates are $\bm{q} = \alpha$ and the remaining terms of Eq. (\ref{eq:modeln}) are defined as follows:
\begin{align}
    & M = m r^2 + I  \\
    & k = 0 \\
    & h = m r g \sin(\alpha)
\end{align}
where $m = 5$\,kg, $r = 0.35$\,m, $l = 0.7$\,m, and $I = 0.051$\,kg.m$^2$ denote, respectively, the mass, center of mass, length, and moment of inertia about the \(Z\)-axis of the robotic arm. The parameter $g = 9.81$\,m/s$^2$ denotes the gravitational acceleration.

The validation was performed using a digital twin of the Quanser QArm within the Quanser Interactive Labs environment. This setup allows for the precise emulation of the physical arm's dynamics, including joint friction and actuator constraints. Each experimental trial was conducted for a duration of 62 seconds at a sampling frequency of 500 Hz using hardware-timed acquisition. Before each experiment, the manipulator was driven to a predefined home configuration and allowed to settle to ensure identical initial conditions. Three distinct reference trajectories were evaluated: square, triangular, and sinusoidal waveforms.

\subsection{Performance Metrics}

Tracking performance was quantified using the root-mean-square (RMS) tracking error
\begin{equation}
\mathrm{RMS}_e = \sqrt{\frac{1}{T}\int_{0}^{T} e^2(t)\,dt}, \qquad e(t)=q(t)-q_d(t),
\end{equation}
and the integral of absolute error (IAE)
\begin{equation}
\mathrm{IAE} = \int_{0}^{T}\lvert e(t)\rvert\,dt.
\end{equation}
In this work, $q(t)$ and $q_d(t)$ are expressed in degrees, hence $\mathrm{RMS}_e$ is reported in degrees (deg) and $\mathrm{IAE}$ in degree-seconds (deg$\cdot$s).

Control effort was quantified using the RMS of the commanded PWM signal
\begin{equation}
\mathrm{RMS}_{\mathrm{PWM}} = \sqrt{\frac{1}{T}\int_{0}^{T}\mathrm{PWM}^2(t)\,dt},
\end{equation}
where PWM is bounded by the actuator saturation limits.

For the step-like (square) reference, transient performance was additionally characterized by overshoot and settling time. Overshoot is reported as a percentage of the step amplitude, and settling time is defined as the time required for $\lvert e(t)\rvert$ to enter and remain within a $\pm 2\%$ band around the commanded step for the specified evaluation window.

\subsection{Sinusoidal Tracking}

\begin{figure}[ht]
    \centering
    \includegraphics[width=\linewidth]{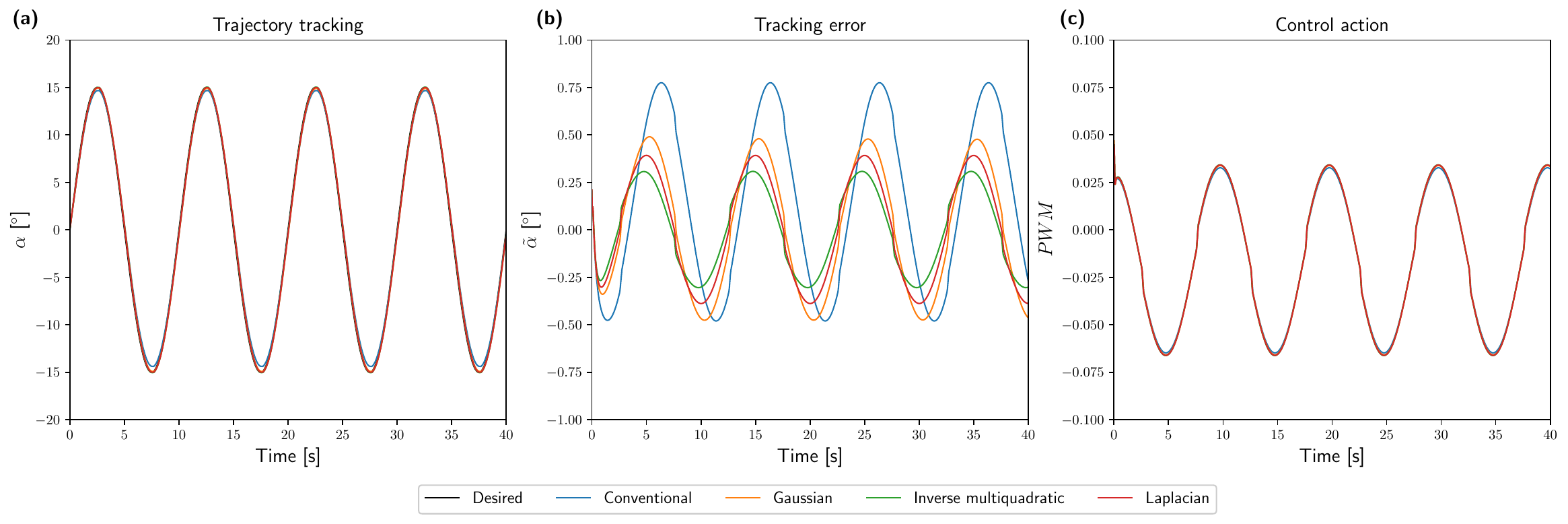}
    \caption{Results for the sinusoidal function.}
    \label{fig:sine}
\end{figure}

The sinusoidal reference, Fig.~\ref{fig:sine}, test evaluates steady-state tracking under smooth excitation. Performance is compared between the baseline controller (conventional feedback linearization controller, $\eta=0$) and the intelligent controller using Gaussian, Laplacian, and inverse multiquadratic (IMQ) activation kernels under identical tuning parameters. Quantitative results are reported in Table~\ref{tab:sine_metrics}.

\subsection{Step-like Tracking (Square Reference)}

\begin{figure}[ht]
    \centering
    \includegraphics[width=\linewidth]{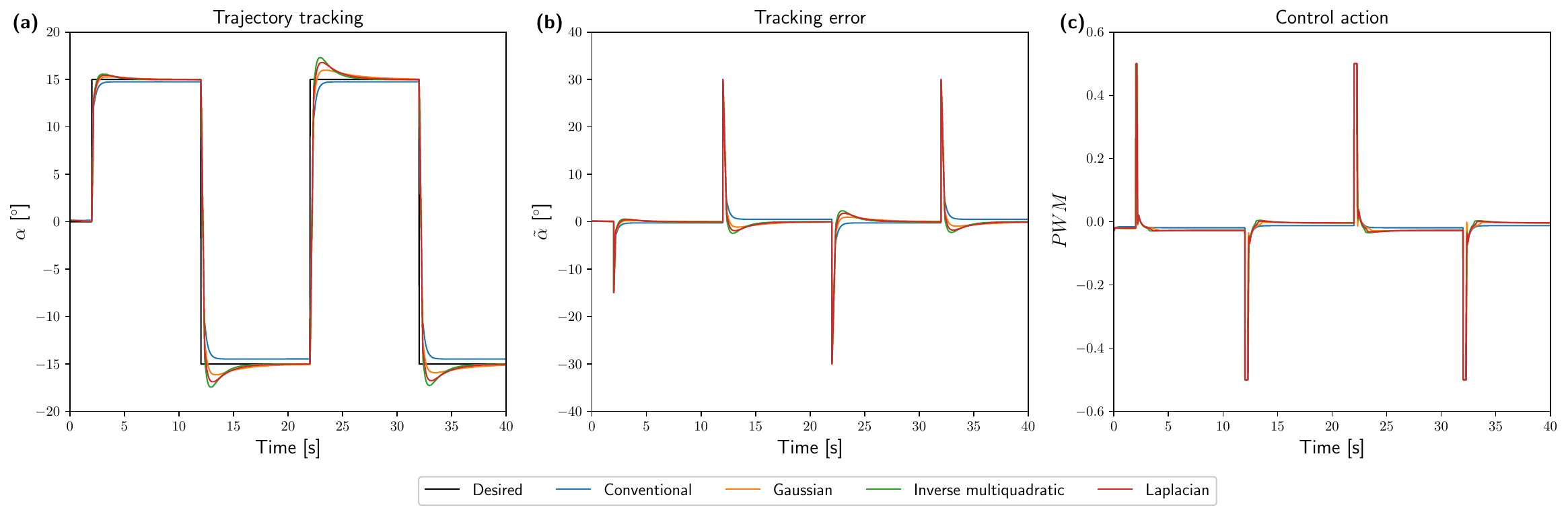}
    \caption{Results for the square function.}
    \label{fig:square}
\end{figure}

The square reference, Fig.~\ref{fig:square}, test evaluates transient behavior under discontinuous commands. In addition to $\mathrm{RMS}_e$, $\mathrm{IAE}$, and $\mathrm{RMS}_{\mathrm{PWM}}$, overshoot and settling time are reported for each kernel. Quantitative results are reported in Table~\ref{tab:square_metrics}.

\subsection{Ramp-like Tracking (Triangular Reference)}

\begin{figure}[ht]
    \centering
    \includegraphics[width=\linewidth]{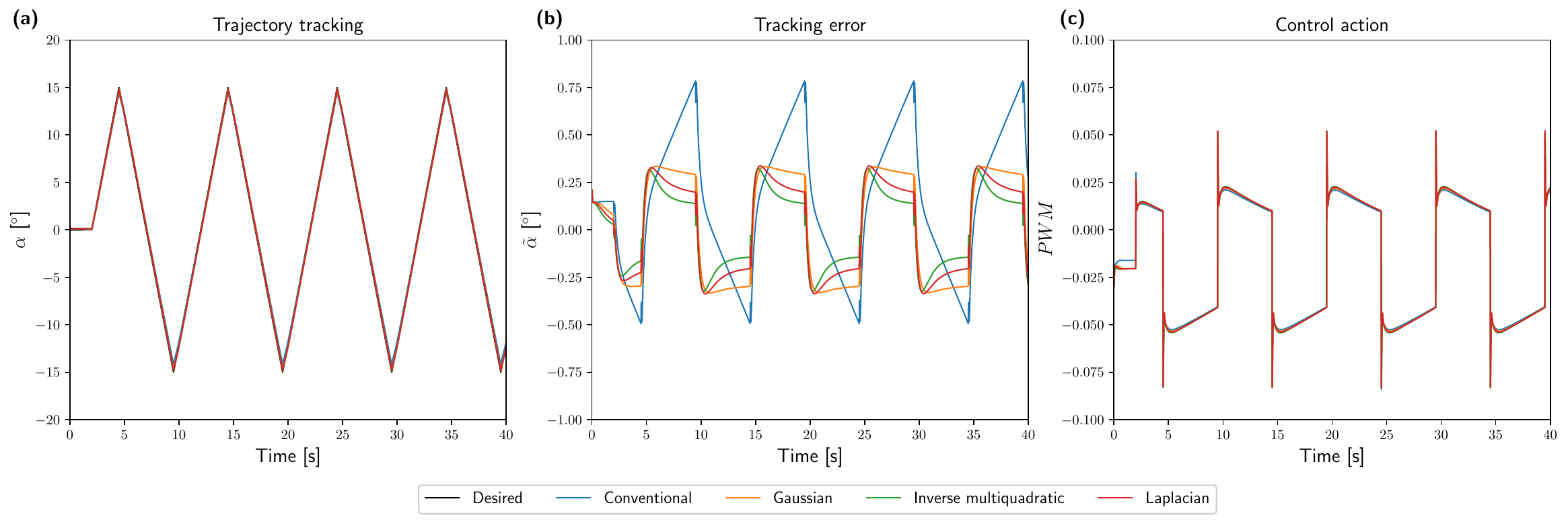}
    \caption{Results for the triangular function.}
    \label{fig:triangle}
\end{figure}

The triangular reference, Fig.~\ref{fig:triangle}, test evaluates tracking over constant-velocity segments (ramp-like motion). Quantitative results are reported in Table~\ref{tab:triangle_metrics}.

\subsection{Quantitative Comparison}

Tables~\ref{tab:sine_metrics}--\ref{tab:triangle_metrics} summarize the mean tracking metrics for each reference type. For the square reference, overshoot and settling time are included to capture transient response characteristics.

\begin{table}[h!]
\centering
\caption{Sine reference metrics (degrees).}
\begin{tabular}{|c|c|c|c|c|}
\hline
Kernel & RMS$_e$ [deg] & Imp. [\%] & IAE [deg$\cdot$s] & RMS$_{\mathrm{PWM}}$ [-] \\
\hline
Baseline ($\eta=0$) & 0.464 & ---    & 25.18 & 0.0381 \\
Gaussian            & 0.341 & 26.5\% & 19.38 & 0.0388 \\
Laplacian           & 0.282 & 39.2\% & 16.02 & 0.0390 \\
IMQ                 & 0.223 & 51.9\% & 12.65 & 0.0391 \\
\hline
\end{tabular}
\label{tab:sine_metrics}
\end{table}

\begin{table}[h!]
\centering
\caption{Square reference metrics (degrees; transient metrics included).}
\begin{tabular}{|c|c|c|c|c|c|c|}
\hline
Kernel & RMS$_e$ [deg] & IAE [deg$\cdot$s] & RMS$_{\mathrm{PWM}}$ [-] & Overshoot [\%] & Settling [s] & sat\_frac [-] \\
\hline
Baseline ($\eta=0$) & 3.095 & 57.76 & 0.0776 & 0.00 & 1.277 & 0.0213 \\
Gaussian            & 3.106 & 49.09 & 0.0829 & 2.85 & 2.567 & 0.0248 \\
Laplacian           & 3.127 & 50.92 & 0.0837 & 5.32 & 2.919 & 0.0249 \\
IMQ                 & 3.142 & 50.45 & 0.0852 & 6.86 & 2.624 & 0.0261 \\
\hline
\end{tabular}
\label{tab:square_metrics}
\end{table}

\begin{table}[h!]
\centering
\caption{Triangle reference metrics (degrees).}
\begin{tabular}{|c|c|c|c|c|}
\hline
Kernel & RMS$_e$ [deg] & Imp. [\%] & IAE [deg$\cdot$s] & RMS$_{\mathrm{PWM}}$ [-] \\
\hline
Baseline ($\eta=0$) & 0.398 & ---    & 20.86 & 0.0356 \\
Gaussian            & 0.292 & 26.6\% & 17.65 & 0.0362 \\
Laplacian           & 0.247 & 37.9\% & 14.83 & 0.0364 \\
IMQ                 & 0.203 & 49.0\% & 11.89 & 0.0365 \\
\hline
\end{tabular}
\label{tab:triangle_metrics}
\end{table}

\subsection{Discussion}

The results demonstrate that activation function selection systematically influences the trade-off between tracking precision and transient aggressiveness, even though all kernels preserve boundedness and ultimate convergence under the same Lyapunov-based adaptation law. A key finding is the superior performance of the inverse multiquadratic (IMQ) kernel for continuous trajectories; it achieved a 51.9\% reduction in RMS error for sine waves and a 49.0\% reduction for triangle waves compared to the baseline. This suggests that kernels with global support, such as the IMQ function, provide a more holistic and accurate approximation of the disturbance manifold when the reference is smooth by enabling wider generalization of the estimate.

In contrast, the square reference tests (Table 2) reveal a critical design trade-off. While the neural kernels improved the IAE (reducing it by up to 15.0\% for the Gaussian case), they introduced transient overshoot (up to 6.86\% for IMQ) and increased settling times. This indicates that more localized kernels lead to sharper adaptation around specific operating regions, which may increase transient aggressiveness under discontinuous commands. These observations indicate that kernel support and smoothness characteristics directly affect the evolution of the adaptive weights and the effective disturbance compensation, suggesting that localized kernels or the baseline controller may be preferable for applications requiring strictly monotonic step responses.

\section{CONCLUDING REMARKS} 

This paper investigated the influence of radial basis activation functions on the closed-loop dynamics of an intelligent controller for robotic manipulators. The proposed framework combines model-based nonlinear control with an RBF neural disturbance estimator updated through a Lyapunov-based adaptation law, guaranteeing boundedness of all closed-loop signals and ultimate convergence of the tracking error to a compact region.

A systematic experimental comparison was conducted on the Quanser QArm platform (Quanser Interactive Labs) under sinusoidal, square, and triangular reference trajectories. The results demonstrate that, although stability properties are preserved for all bounded kernels, activation function selection significantly influences transient response and steady-state accuracy while maintaining a consistent control effort. The most significant result is that the IMQ kernel can reduce tracking error by approximately 52\% while the control effort ($\text{RMS}_{\text{PWM}}$) increased by less than 3\% relative to the baseline.

In particular, the inverse multiquadratic kernel exhibited improved steady-state tracking performance for smooth and ramp-like trajectories, while more localized kernels led to sharper adaptation and increased transient aggressiveness under discontinuous commands. These findings indicate that kernel support and smoothness characteristics directly affect the disturbance approximation quality and the evolution of the adaptive weights.

Overall, the study highlights that activation function choice should be regarded as a structural design parameter in intelligent controllers that allows engineers to tune for either maximum steady-state precision or minimal transient overshoot. Future work will extend the analysis to multi-DOF configurations and investigate systematic kernel tuning strategies that balance approximation capability and transient robustness.

\section{REFERENCES} 

\bibliographystyle{eurodiname2026}
\renewcommand{\refname}{}
\bibliography{bibfile}

\section{RESPONSIBILITY NOTICE}

The authors are solely responsible for the printed material included in this paper.

\end{document}